\documentclass[lettersize,journal]{IEEEtran}
\usepackage{amsmath,amsfonts}
\usepackage{algorithmic}
\usepackage{algorithm}
\usepackage{array}
\usepackage[caption=false,font=normalsize,labelfont=sf,textfont=sf]{subfig}
\usepackage{textcomp}
\usepackage{stfloats}
\usepackage{url}
\usepackage{verbatim}
\usepackage{soul}
\usepackage{graphicx}
\usepackage{cite}
 \usepackage{soul, color, xcolor} 
\usepackage{orcidlink}
\usepackage[marginal]{footmisc}

\begin{document}







\title{AGCo-MATA: Air-Ground Collaborative Multi-Agent Task Allocation in Mobile Crowdsensing}
\author{ Tianhao Shao, Bohan Feng, Yingying Zhou, Bin Guo, and Kaixing Zhao }

\maketitle
\begin{abstract}
Rapid progress in intelligent unmanned systems has presented new opportunities for mobile crowd sensing (MCS). Today, heterogeneous air-ground collaborative multi-agent framework, which comprise unmanned aerial vehicles (UAVs) and unmanned ground vehicles (UGVs), have presented superior flexibility and efficiency compared to traditional homogeneous frameworks in complex sensing tasks. Within this context, task allocation among different agents always play an important role in improving overall MCS quality. In order to better allocate tasks among heterogeneous collaborative agents, in this paper, we investigated two representative complex multi-agent task allocation scenarios with dual optimization objectives: (1) For AG-FAMT (Air-Ground Few Agents More Tasks) scenario, the objectives are to maximize the task completion while minimizing the total travel distance; (2) For AG-MAFT (Air-Ground More Agents Few Tasks) scenario, where the agents are allocated based on their locations, has the optimization objectives of minimizing the total travel distance while reducing travel time cost. To achieve this, we proposed a Multi-Task Minimum Cost Maximum Flow (MT-MCMF) optimization algorithm tailored for AG-FAMT, along with a multi-objective optimization algorithm called W-ILP designed for AG-MAFT, with a particular focus on optimizing the charging path planning of UAVs. Our experiments based on a large-scale real-world dataset demonstrated that the proposed two algorithms both outperform baseline approaches under varying experimental settings, including task quantity, task difficulty, and task distribution, providing a novel way to improve the overall quality of mobile crowdsensing tasks.
\end{abstract}

\begin{IEEEkeywords}
mobile crowdsensing, unmanned aerial vehicle, unmanned ground vehicle, multi-agent framework, multi-task allocation.
\end{IEEEkeywords}

\section{Introduction}
\IEEEPARstart{I}{n} the era of Internet of Things, mobile crowdsensing plays an important role in improving the efficiency and quality of social governance\cite{ref-1}. For example, mobile crowdsensing can provide essential data for cost accounting in city construction tasks \cite{mednis2011real}. In addition, as an emerging field, it also involves various application scenarios, including real-time traffic monitoring, precipitation monitoring, and air quality assessment \cite{mednis2011real, mohan2008nericell}. In recent years, although mobile devices such as smartphones are becoming common in many mobile crowdsensing tasks \cite{ref-2}, their large-scale application still faces many challenges, especially when dealing with scenarios where human presence is difficult or impossible. In this case, heterogeneous multi-agent framework, especially air-ground robot systems such as unmanned aerial vehicles (UAVs) and unmanned ground vehicles (UGVs) can collaboratively play a significant role in conducting real-time complex sensing tasks \cite{wang2019development}.

Although scenarios involving a single agent (for example, a moving camera based on an unmanned aerial vehicle) have been extensively studied before \cite{wang2019development}, their applications in more challenging scenarios are still limited due to restricted capabilities \cite{xu2020learning, gao2022uav}. As unmanned technology rapidly advances, UAVs and UGVs are increasingly utilized collaboratively in numerous fields. Such convergence of technologies has given rise to heterogeneous multi-agent framework, particularly the air-ground collaborative framework focused in this paper. Within such framework, UAVs with their aerial mobility and rapid coverage capabilities, can collaborate with UGVs, which possess greater load capacity and longer operational durations \cite{xi2024lightweight}, \cite{yu2014cooperative}, to perform complex mobile sensing tasks. Compared to homogeneous robot systems, heterogeneous multi-agent framework exhibit distinct advantages in flexibility, task allocation efficiency, and task diversity, making them a valuable research direction for mobile crowdsensing \cite{ref-3}.

At the same time, both UAVs and UGVs face challenges in performing multiple missions. Although UAVs are highly maneuverable and cover a wide range of areas, they are limited by battery life, restricting their flight endurance and sensing range \cite{zeng2017energy}. In terms of UGVs, although they can offer longer endurance and greater autonomy, they generally move slower and have a more limited sensing coverage. In this case, combining UAVs and UGVs into air-ground collaborative frameworks may tackle these limitations and achieve complementary advantages. Such collaboration can not only balance the sensing coverage and sensing time, but may also improve the overall efficiency of mobile crowdsensing tasks \cite{wang2023air, ding2021review}. In addition, moving from different angles and altitudes during air and ground collaborations can also improve the comprehensiveness of sensing and create more opportunities for effective sensing task execution \cite{sun2021aoi, wei2022high}.

However, although various air-ground collaborative frameworks have been studied in recent years \cite{dai2022aoi, ye2023qoi}, as one of the most critical mechanisms, heterogeneous task allocation has been less investigated before \cite{ref-4}, especially for the collaborative UAVs-UGVs framework. Traditional task allocation methods are designed mainly for homogeneous systems, and agents are usually selected based on uniform criteria \cite{ref-5}. But for air-ground collaborative frameworks, its heterogeneity complicates task allocation and requires more careful consideration of differences between agents \cite{ref-5}. In fact, for ubiquitous systems (i.e. air-ground collaborative framework in our case) which always have limited computing resources, reasonable task allocation can not only improve the overall task completion efficiency, optimize the agent movement paths, but can also greatly reduce the redundant movements of different agents \cite{ref-4}.

\begin{figure}[htbp]
    \centering
    \includegraphics[width=1\linewidth]{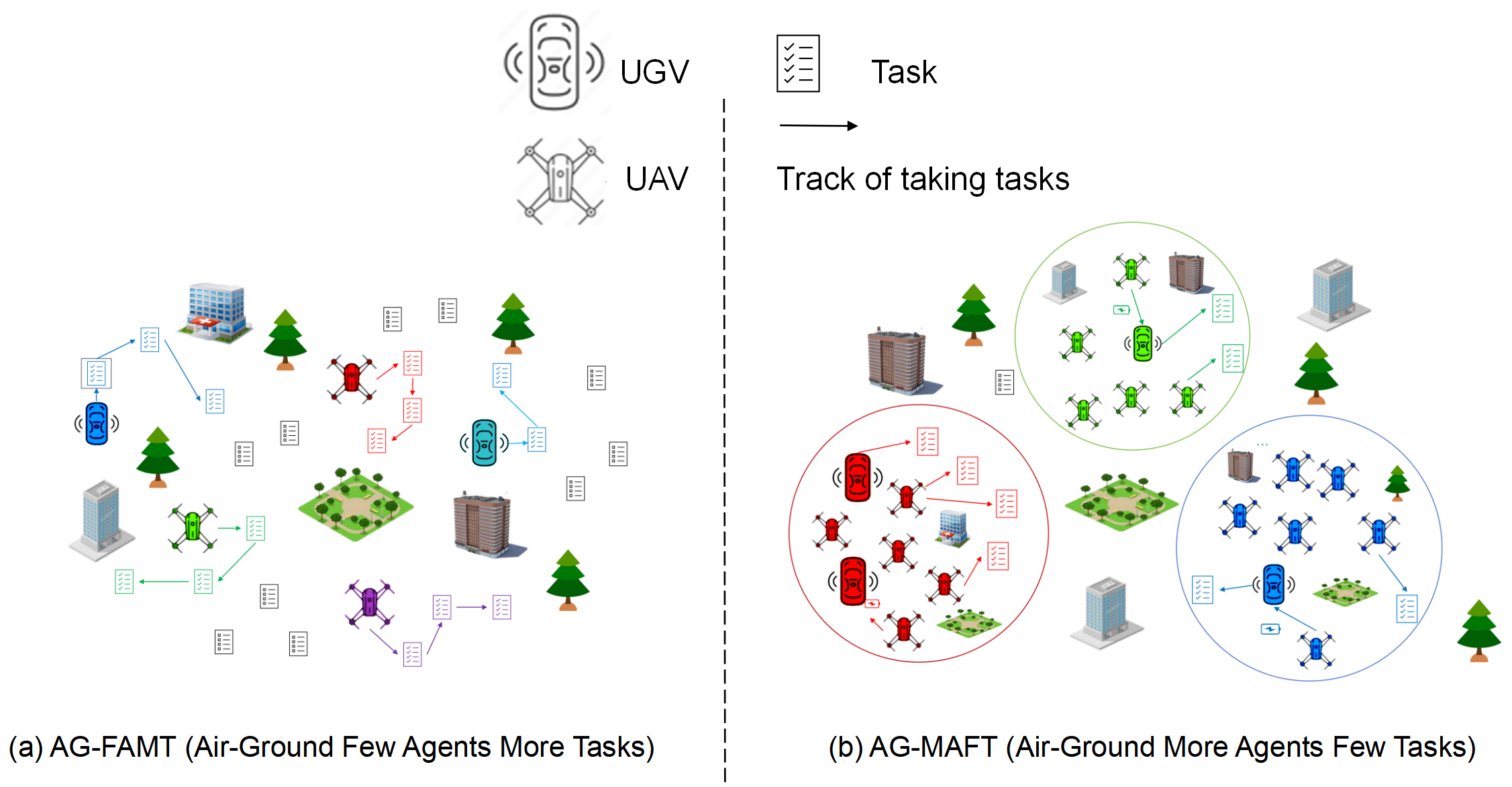}
    \vspace{-0.5em}
    \caption{Two scenarios of agents selection.}
    \label{fig1}
\end{figure}

To systematically address these challenges, in this paper, we focus on two typical task-agent configurations \cite{liu2016taskme}: the few agents more tasks and the more agents few tasks scenarios, which we call AG-FAMT (Air-Ground Few Agents More Tasks) and AG-MAFT (Air-Ground More Agents Few Tasks) (as shown in Figure \ref{fig1}). In the AG-FAMT scenario as shown in Figure \ref{fig1}(a), the objectives are to maximize task completion while minimizing the total travel distance. In the AG-MAFT scenario as shown in Figure \ref{fig1}(b), privacy preservation is prioritized and the appropriate agent is selected to perform the task based on the proximate location, with the objectives of minimizing the total travel distance while reducing the cost of travel time. In general, this study aims to improve the efficiency of the assignment of tasks in heterogeneous collaborative air-ground frameworks when performing mobile crowdsensing tasks \cite{zhao2014coupon,zhao2015opportunistic,zhu2018spatiotemporal}.

The main contributions of this paper are as follows:

1. We designed a heterogeneous task allocation framework for air-ground collaborative framework consisting of UAVs and UGVs, which leveraged their complementary characteristics to improve the efficiency of task completion. By considering factors such as endurance time, movement speed, travel distance, and sensing range, our framework can achieve adapted sensing task allocation.

2. We proposed a bi-objective optimization method for heterogeneous task allocation and designed the corresponding allocation mechanisms. For AG-FAMT, we aim to maximize the number of task completions and minimize the total travel distance; For AG-MAFT, we ensure task execution while minimizing the total travel distance and the travel time cost. In addition, we implemented optimized charging paths for the UAVs to support sustained mission execution and efficient use of resources.

3. Through a series of experiments, we verified the effectiveness and superiority of our proposed framework, which provides application potential for future mobile crowdsensing.

The following sections will present the design details and implementation of our framework. Section 2 reviews the related works. The system model and problem formulation based on the AG-FAMT and AG-MAFT scenarios are introduced in Section 3. Section 4 provides the performance evaluation. The paper is closed with the evaluation and conclusion in Section 5 and 6.

\section{Related Work}
\subsection{Mobile Crowdsensing}
Mobile Crowdsensing (MCS) is a technology that uses widely distributed mobile devices to collect spatiotemporal data. It integrates the perception, computing and communication capabilities of mobile devices to achieve monitoring and data collection in multiple fields such as the environment and society \cite{liu2015energy}, \cite{zhang2017learning}. In recent years, mobile crowd sensing technology has received widespread attention and research.Mohan et al. \cite{mohan2008nericell} demonstrated the feasibility of continuous environmental monitoring through the use of low-cost mobile devices. Similarly, Mednis et al. \cite{mednis2011real} focused on real-time monitoring of road surfaces to enhance urban surveillance capabilities. Additionally, Pawar et al. \cite{pawar2020efficient} proposed an integration of mobile sensors with neural networks, which significantly improved the accuracy of real-time urban sensing systems.


Existing MCS platforms primarily focus on three aspects: task publishing \cite{9475480}, data collection \cite{10050805} as well as participant selection \cite{8254369}, and usually concentrate on single-task scenarios \cite{capponi2019survey}. Meanwhile, with the advancement of MCS technology, the task allocation problem has emerged as a significant challenge that impacts the efficiency and quality of MCS operations \cite{10050805}. To address the privacy issues in mobile crowdsensing, Liu et al. ensured the accuracy and integrity of the data by reducing the communication volume between aggregation nodes, and this approach has stronger privacy protection capabilities compared with traditional sensing auction models.



\subsection{Air-Ground Collaborative Systems}
As Mobile Crowdsensing (MCS) technology encounters increasing challenges, meeting task requirements in complex environments becomes difficult due to the uniform capabilities of agent and homogeneous robotic systems. To overcome these limitations, the development of multi-agent technology and the widespread application of multi-agent frameworks across various fields have prompted the exploration of air-ground collaborative frameworks. This approach integrates unmanned vehicles, such as unmanned aerial vehicles (UAVs) and unmanned ground vehicles (UGVs), to enhance MCS efficiency and improve the accuracy of data acquisition and mapping. Specifically, UAVs are employed to collect large-scale data, while UGVs provide auxiliary ground support, thereby enhancing perception efficiency \cite{wang2022task}, \cite{he2020ground}. For example, Sun et al. \cite{sun2021aoi} proposed a UAV-based data collection method for Internet of Things (IoT) networks that optimizes flight speed, hovering position, and bandwidth allocation. Meanwhile, Wei et al. \cite{wei2022high} improved UAV navigation and mission efficiency through an incentive mechanism for path planning in complex environments.

Meanwhile, the field of task allocation within multi-agent frameworks continues to confront several challenges. As the scale of these frameworks expands and heterogeneous intelligences proliferate, traditional centralized task assignment methods face significant obstacles related to computational complexity and communication overhead. Distributed task allocation algorithms have emerged as a common approach in heterogeneous systems, enabling each agent to perform task allocation based on its own state and task requirements. This approach alleviates the computational burden associated with centralized scheduling and enhances system scalability \cite{yu2021proficiency}, \cite{gao2023asymmetric}. Additionally, task allocation methods grounded in game theory have garnered substantial attention in the literature \cite{zhang2018dynamics}, \cite{cui2013game}. Despite these advancements, human intervention remains essential for providing real-time feedback, managing anomalies, and validating data \cite{wang2022human}, \cite{ding2021crowdsourcing}.

\subsection{Task Allocation}
Task allocation \cite{guo2018task} involves assigning tasks to appropriate executors, which is crucial in Mobile Crowdsensing (MCS) as the selected executors significantly influence the efficiency and quality of task performance. Generally, there are two primary approaches to task assignment in crowdsourcing: 1) autonomous selection by executors \cite{ra2012medusa} and 2) allocation by a central server \cite{kazemi2012geocrowd}. Liu et al. \cite{liu2016taskme} investigated the impact of various factors on multi-task assignment within mobile crowds, analyzing three key aspects: the number of participants, the number of tasks, and task urgency. In another study, Song et al. \cite{song2014qoi} addressed the diverse quality of information (QoI) requirements of spatial crowdsourcing tasks and proposed a multi-task allocation strategy aimed at selecting a minimal subset of workers to satisfy the QoI requirements of concurrent tasks while adhering to total budget constraints. Additionally, ActiveCrowd \cite{guo2016activecrowd} explored the multitask workload scheduling problem under both intentional and unintentional movement scenarios, thereby catering to time-sensitive and delay-tolerant tasks, respectively.

The preceding analysis clearly highlights the numerous advantages of air-ground collaborative frameworks. In this paper, we propose a multi-task allocation framework that integrates UAVs and UGVs, focusing on the relationship between the number of tasks and the number of agents. We investigate a universal task allocation method that is informed by a comprehensive understanding of the fundamental constraints associated with task allocation, employing a comparative analysis of task and agent quantities.

\section{SYSTEM MODEL AND PROBLEM FORMULATION}

In this section, we introduce the system model and problem description.

\subsection{System Model}

As shown in Figure \ref{fig1}, we consider a heterogeneous multi-agent air-ground collaborative MCS framework involving obstacles (e.g. buildings), unmanned aerial vehicles (UAVs), unmanned ground vehicles (UGVs), task points, and boundary areas. Let the set of UAVs be denoted as $X = \{{x_1, x_2, ..., x_s}\}$, the set of UGVs as $Y = \{{y_1, y_2, ..., y_t}\}$, and the overall agent set as $U = \{{X + Y}\} = \{{x_1, x_2, ..., x_s, y_1, y_2, ..., y_t}\}$. The task set is defined as $T = \{{t_1, t_2, ..., t_n}\}$, where the number of UAVs is $s$, the number of UGVs is $t$, the number of tasks is $N$ and the total number of agents $M = s + t$. Obstacles are represented as $Ob = \{{ob_1, ob_2, ..., ob_a}\}$, with a total of $a$ obstacles. The agents and tasks are distributed within a physically bounded area.

The basic information for each task can be expressed as a tuple $<(x_i,y_i,h_i),c_{i1},c_{i2}, ... , c_{in}>$, with UAV information given by $<(x_j,y_j,h_j),c_{j1},c_{j2}, ... , c_{jn}>$, and UGV information as $<(x_k,y_k,h_k),c_{k1},c_{k2}, ... , c_{kn}>$. This information includes location details, represented in a 3D Cartesian coordinate system $(x, y, h)$. 
The set $\{c_{i1}, c_{i2}, \ldots, c_{in}\}$ represents the required capability levels across types for task $t_i$ $(1, 2, \ldots, n)$, while $\{c_{j1}, c_{j2}, \ldots, c_{jn}\}$ and $\{c_{k1}, c_{k2}, \ldots, c_{kn}\}$ respectively denote the capability levels of UAVs and UGVs for each capability type $(1, 2, \ldots, n)$.

\subsection{Problem Description}
\subsubsection{AG-FAMT (Air-Ground Few Agents More Tasks) ($N > M$)}

In the AG-FAMT scenario, the number of tasks exceeds the number of agents, often involving urgent tasks, such as post-storm urban environment sensing (e.g., monitoring water levels and assessing damage to buildings and trees after a typhoon). Each agent needs to complete multiple tasks and move to the designated locations to accomplish these tasks, making travel distance a significant factor: shorter travel distances reduce task completion time.

The optimization objectives in this case are to maximize the total number of completed tasks and minimize the total travel distance of the agents. Due to the urgency of tasks and limited agent resources, incentive costs and the risk of precise location information leakage of the agents are ignored in this scenario. To enhance the quality of sensing tasks, multiple agents are required to collaborate on each task, with an upper limit on the number of agents allowed for each task to avoid data redundancy. In the same way, in AG-FAMT, each agent needs to complete multiple tasks. In order to balance task quality and agent fairness, a task upper limit of $q_i,{(i=1,2)}$ is set for each type of agent, where $q_1$ is the task upper limit of UAV and $q_2$ is the task upper limit of UGV. Given the urgency of the task, precise location information of agents becomes essential, so privacy concerns are not considered in this case.
We assume that the travel time is proportional to the travel distance, thus transforming the goal of minimizing the task completion time into the goal of minimizing the travel distance.

\subsubsection{AG-MAFT (Air-Ground More Agents Few Tasks) ($N < M$)}

In the scenario of AG-MAFT, as opposed to the case of AG-FAMT, there are more agents than tasks. More precisely, in this scenario, some tasks do not require extensive involvement of a large number of agents while some other tasks require continuous execution, such as air quality monitoring and continuous urban safety patrols. For AG-MAFT, with sufficient agent resources, agents can be flexibly assigned to improve the quality of tasks. Here, each agent is required to complete only one task and an initial check is performed to ensure that they meet the minimum competency standards to perform the task. Due to the non-urgent nature of the tasks, precise location data of the agents is not required, and tasks can be selected within a defined range around each agent's approximate location to maintain privacy.


This study classifies tasks according to distance and spatial requirements. Given that UAVs can operate at high speeds, they can greatly expand the perception scope of long-distance tasks. In contrast, UGVs are more suitable for close-range tasks, which can reduce the travel distance and improve the response efficiency. To ensure the successful completion of tasks, it is essential to optimize task allocation and select agents that meet the minimum threshold requirements for successful execution. In the scenario of AG-FAMT, we strive to maximize the number of completed tasks while minimizing the total travel distance. In the AG-MAFT scenario, we aim to complete all tasks while minimizing the agents' travel time cost and the total travel distance. Moreover, considering the continuity of tasks, this paper has specifically designed a path planning mechanism for UGVs that are responsible for providing charging support for UAVs. Another key objective of this path planning mechanism is to ensure that the total travel distance is not excessive, especially in the scenario of continuous MCS.  

\subsection{Problem Formulation}

\subsubsection{AG-FAMT Formulation}

Within a defined area, given a set of agents $U = {X + Y} = \{{x_1, x_2, ..., x_s, y_1, y_2, ..., y_t}\}$, assume that the task execution limits for heterogeneous robots are $q = {q_1, q_2}$ (where $q_1$ is the task limit for UAVs and $q_2$ is the task limit for UGVs). The set of published tasks is $T = \{{t_1, t_2, ..., t_n}\}$, where each task $t_i$ can be assigned to at most $p_i$ agents (assignment to UAVs is represented by $x_{ij}$, and assignment to UGVs by $y_{ik}$). We use $UT_i = \{{x_{i1}, x_{i2}, ..., y_{i1}, y_{i2}}\}$ to denote the set of agents assigned to task $t_i$, and $TU_j = \{{x_{1j}, x_{2j},...}\}$ or $TU_k = \{{y_{1k}, y_{2k},...}\}$ to respectively represent the set of tasks assigned to UAV $x_j$ and UGV $y_k$. Let $S$ denote the maximum movable distance for an agent. Considering the endurance limitations of agents, the travel distance required to execute each task must be evaluated before task assignment. Additionally, $D(TU_j)$ and $D(TU_k)$ represent the total travel distance for completing the set of tasks assigned to each UAV and UGV, respectively. Therefore, AG-FAMT problem can be formulated using Eq. \ref{equ-01} to Eq. \ref{equ-06}:

\begin{equation}
\begin{split}
\sum_{i=1}^q & \sqrt{(x_j \ - \ x_i)^2 \ + \ (y_j \ - \ y_i)^2 \ + \ (h_j \ - \ h_i)^2} \\
& = D(TU_{j/k})
\end{split}
\label{equ-01}
\end{equation}

\begin{equation}
maximize \sum_{i=1}^m |TU_{j/k}|
\label{equ-02}
\end{equation}
\begin{equation}
minimize \sum_{i=1}^m D(TU_{j/k})
\label{equ-03}
\end{equation}
\begin{equation}
|TU_{j/k}| = q \ (q= q_1,q_2 ),(1 \leq j \leq s,1 \leq k \leq t)
\label{equ-04}
\end{equation}
\begin{equation}
|UT_i| <= p_i \ (1 \leq i \leq n)
\label{equ-05}
\end{equation}
\begin{equation}
D(TU_{j/k}) \ \leq \ S
\label{equ-06}
\end{equation}

The AG-FAMT problem presents two primary challenges: first, ensuring that each agent can complete multiple tasks, and second, achieving the dual optimization objectives of maximizing the total number of completed tasks and minimizing the agents' total travel distance. Addressing the first challenge involves selecting multiple tasks for each agent, which is a complex dynamic process. In the AG-FAMT scenario, not only is there a combinatorial problem between tasks and agents, but the execution path for each agent must also be planned to achieve the shortest travel distance. Given the combination of task allocation and path optimization, finding a comprehensive algorithm to solve the AG-FAMT problem is challenging. Therefore, we plan to calculate the set of all executable tasks for each agent and seek an optimal solution to the combinatorial optimization problem, which could be an effective approach. For the second challenge, achieving both optimization objectives simultaneously is infeasible, as they are often contradictory. We will balance the objectives of maximizing task completion and minimizing total travel distance to find a compromise solution.

On the basis of analyzing the AG-FAMT problem, we adopt the Minimum Cost Maximum Flow (MCMF) model for solving it \cite{hadji2012minimum}. The MCMF model is a classic method for solving dual objective optimization problems. We transformed the AG-MAFT problem into an MCMF problem and constructed a new MCMF model called MT-MCMF by setting various constraints to achieve effective task allocation and path planning. This method will help ensure the completion of tasks while minimizing the total travel distance of agents, thereby optimizing overall system performance.

\subsubsection{MT-MCMF Algorithm}

The MT-MCMF (Multi Task Minimum Cost Maximum Flow) algorithm is based on MCMF theory, with the goal of maximizing the number of completed tasks while minimizing the total travel distance of a given set of agents and tasks. The core of MCMF model is to find the optimal path set that generates the maximum traffic at the minimum cost. We transform the AG-MAFT problem into an MCMF problem, where the total travel distance represents cost and the total number of completed tasks represents flow. We combine the minimum cost and maximum flow problem with the goals of minimizing travel distance and maximizing task completion.

In the MT-MCMF model graph, each edge has a capacity, flow, and cost. The capacity of an edge denotes the maximum flow it can carry, representing the maximum number of tasks. However, several challenges arise with direct application of the MT-MCMF model: (1) the flow network cannot consist solely of agent and task nodes; (2) representing each agent’s path with $q$ tasks is complex; (3) the distinct requirements of agents and tasks need clear representation within the flow network. To address these issues, we restructured the MT-MCMF model as illustrated in Figure \ref{fig2} and made the following three modifications:

\begin{figure}[htbp]
    \centering
    \includegraphics[width=1\linewidth]{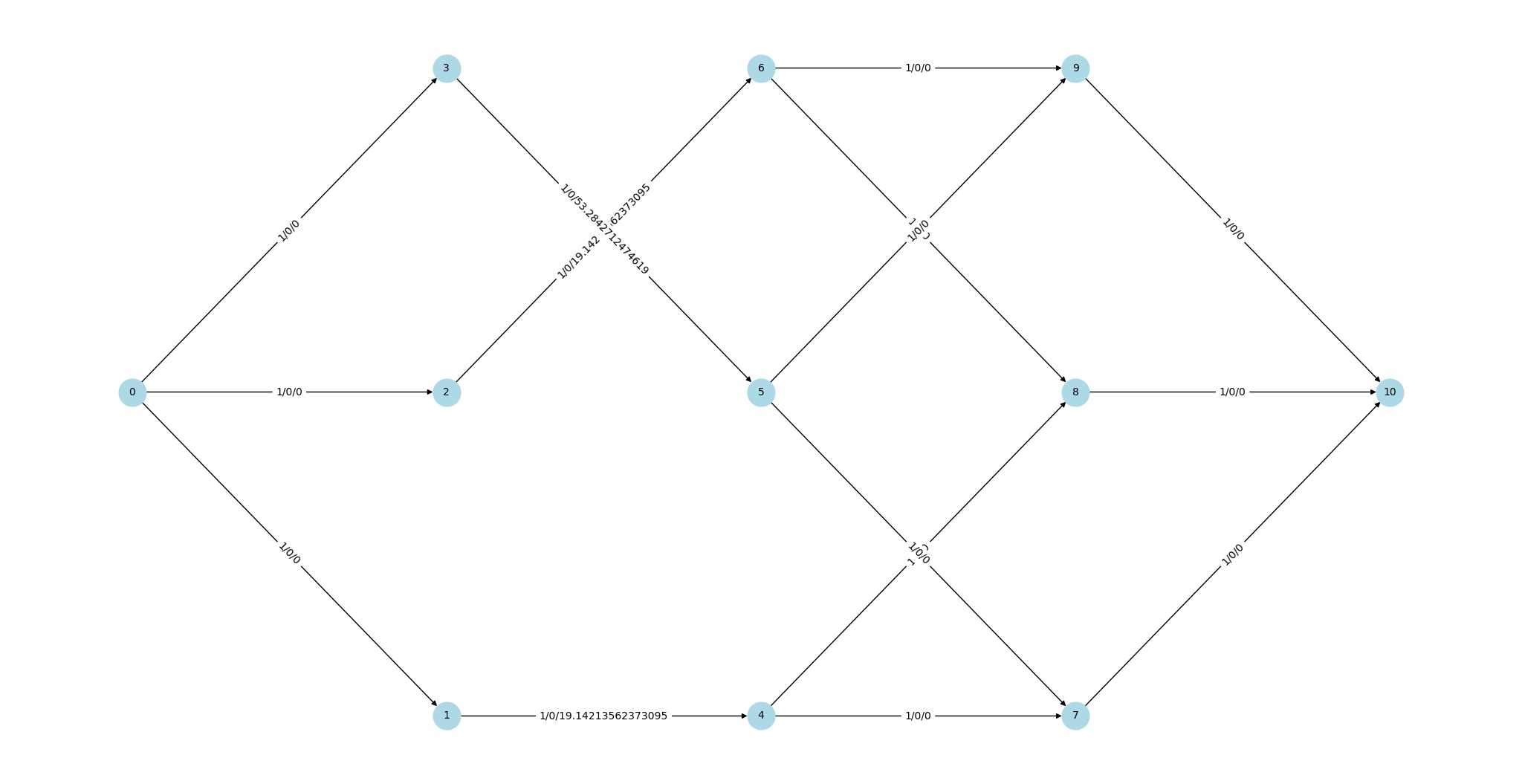}
    \vspace{-0.5em}
    \caption{MT-MCMF model of AG-FAMT problem.}
    \label{fig2}
\end{figure}

(1) Adding source and sink nodes: We introduced source and sink nodes into the flow network. For edges between the source node and agent nodes, each edge has a capacity of $q_i$ $(i = 1,2)$, reflecting the $q_i$ tasks each agent must complete; the cost of these edges is set to zero, as they solely carry incoming flow without involving move distance. Additionally, the flow from agent nodes through task nodes into the sink node reflects task completion. The total maximum flow of the sink node is set to $n \times p$, as each task can be performed by up to $p$ agents. Each edge between task nodes and the sink node has a capacity of $p$ and a cost of zero. The actual flow to the sink node represents the total number of agents completing each task.

(2) Constructing task sets and adding secondary nodes to represent task sets: We pre-enumerate all possible task sets, where each set contains $q$ tasks selected from the total $N$ tasks, yielding a total of $C_N^q$ task sets. Each agent can select any task set to perform. We compute the shortest path between each agent and each task in the task set, assigning this as the cost of the edge. Notably, the task sets in the flow network do not specify task completion order. We draw from the Traveling Salesman Problem (TSP) approach by pre-calculating the shortest paths between each agent and each task within the task set. Consequently, the cost of edges between agent nodes and task set nodes is the $TSP$ distance for the agent to complete tasks in the set.

(3) Adding tertiary nodes to indicate task completion status: We set different capacities on edges to meet the specific requirements of agents and tasks. Assuming that each task set can only be executed by one agent, we set the capacity of edges between task nodes and the sink node to $p$, as each task can be performed by up to $p$ agents.

Through these improvements, we can leverage the MT-MCMF model more effectively to address the AG-MAFT problem, ensuring that the maximum number of tasks is completed while minimizing the total travel distance of agents, thereby optimizing the overall system performance.

\begin{algorithm}[!h]
    \caption{MT-MCMF}
    \renewcommand{\algorithmicrequire}{\textbf{Input:}}
    \renewcommand{\algorithmicensure}{\textbf{Output:}}
    \begin{algorithmic}[1]
    \REQUIRE The agent set \( A(M) \) and the task set \( T(N) \).  
    \ENSURE The selected \( AT \) set and the corresponding minimum distance \( d \).    
    \STATE Connect the source node to each agent node (from \( 0 \) to \( M \)), where the edge cost \( c = 0 \) and the flow upper limit is \( q \).
    \STATE Create \( C(n, q) \) transfer sets \( TT \). Connect each agent node to the task set (from \( M \) to \( M + C_N^q) \)), with edge cost \( c \) equal to the minimum distance (calculated using TSP to obtain the minimum distance).
    \STATE Link each \( TT \) set to each task node, adding connections from \( M + C_N^q \) to \( M + C_N^q + N \), and then connect all these nodes to the sink node.
    \STATE Create the flow network \( (V, E, C, W) \).
    \STATE Initialize the flow \( f = 0 \).
    \WHILE{there exists an augmenting path in the residual network \( G_f \)}
        \STATE Select the augmenting path \( p^* \) with minimum cost.
        \STATE Set \( c_f(p^*) = q \).
        \STATE Augment the flow \( f \) along \( p^* \) with \( c_f(p^*) \).
    \ENDWHILE
    \STATE Return the agent-task assignment \( A \)-\( T \).
    \STATE Output the assignment \( A \)-\( T \) and the total minimum distance.
    \end{algorithmic}
\end{algorithm}

Overall, the MT-MCMF algorithm consists of two main components: constructing the flow network within the MT-MCMF model and finding the optimal solution in the flow network. Given an agent set with $M$ agents and a total of $N$ tasks, each agent is required to complete $q$ tasks. Therefore, these tasks can be grouped into $C_N^q$ task sets, where $C_N^q$ represents the number of combinations of choosing $q$ tasks from $N$ tasks, thus forming the task set $TT$. Consequently, there are a total of $M \times C_N^q$ possible $TSP$ problems for all task-execution sets. Based on our assumptions, the constructed flow network $G = (V, E, C, W)$ includes nodes $V$, edges $E$, capacities $C$ for each edge, and total costs $W$ for each edge. The process of obtaining the optimal solution from the flow network constructed by MT-MCMF is as follows: First, initialize the flow $f$ to zero, then iteratively find an augmenting path until no further paths are available, and return the final flow $f$. Note that the flow of edges between agent nodes and task set nodes can be zero, indicating that the agent did not choose to execute that task set. At the same time, the flow value $q$ means that the agent has carried out q tasks from that specific task set $TT$. Finally, the agent set $Agents$ and the corresponding task set $TT$ that they have accomplished are outputted. Through these steps, the MT-MCMF algorithm effectively optimizes task allocation among agents in the flow network, achieving the goals of maximizing task completion and minimizing total travel distance.

For the AG-FAMT problem, the time complexity of MT-MCMF is $O(m \times C_k^q \times q^3) + O(m \times q \times (m + n) + C_k^q)$. Here, $O(m \times C_k^q \times q^3)$ represents the time complexity for calculating the shortest path using $Christofides$ after constructing the task sets, while $O(m \times q \times (m + n) + C_k^q)$ calculates the time complexity of finding the optimal solution in the flow network for the MT-MCMF model.

\subsubsection{AG-MAFT Formulation}

Given a set of predefined sensing range, represented as $A = \{{A_1, A_2, ..., A_i, ..., A_m}\}$, each zone includes a group of agents regularly active in that area, denoted as $A_i = \{{a_1, a_2, a_3, ...}\}$. Additionally, there are several sensing tasks represented by $T = \{{t_1, t_2, ..., t_j, ... t_n}\}$, where each task $t_j$ is assigned to $p_j$ agents. Note that $D_{ij}$ represents the distance between agents in zone $A_i$ and task $t_j$. Furthermore, $x_{ij}$ denotes the number of agents in zone $A_i$ assigned to complete task $t_j$. Therefore, the AG-MAFT problem can be represented by Eq. \ref{equ-07} to Eq. \ref{equ-10}:
\begin{equation}
minimize \sum_{i=1}^m  \sum_{j=1}^n D_{ij} \times x_{ij}
\label{equ-07}
\end{equation}
\begin{equation}
\sum_{j=1}^n x_{ij} \leq |A_i| \ (i \leq i \leq m) 
\label{equ-08}
\end{equation}
\begin{equation}
\sum_{i=1}^m x_{ij} = p_{ij} (i \leq j \leq n)
\label{equ-09}
\end{equation}
\begin{equation}
x_{ij} \in Z^n 
\label{equ-10}
\end{equation}

Given the ample agent resources under AG-MAFT, we set higher standards for task completion quality, allowing agents to execute tasks only if all capability requirements exceed the task requirements. The capability calculation formulas are shown in Eq. \ref{equ-11} and Eq. \ref{equ-12}:
\begin{equation}
    eff_{ij}\ = \ (\frac{c_{i1}}{c_{j1}} \ + \ \frac{c_{i2}}{c_{j2}} \ + \ ... \ + \ \frac{c_{in}}{c_{jn}} ) \ \times co_{ij}
    \label{equ-11}
\end{equation}
\begin{equation}
    co_{ij} \ = \left\{\begin{array}{ll} 
2 \ , \ if \  \forall \ 1 \leq k \leq n : c_{ik} > c_{jk} \\
\frac{1}{2} \ , \ otherwise
\end{array}\right.
\label{equ-12}
\end{equation}

The main challenge in the AG-MAFT scenario involves two conflicting optimization goals: minimizing movement time costs and travel distances. In order to solve the AG-MAFT problem, we adopt a multi-objective optimization model. Briefly, the main approach to solve a multi-objective optimization problem is to transform it into a single-objective problem. We use linear weighting to achieve this.

\subsubsection{The Linear Weight Method}

The main idea of the linear weight method is to transform a multi-objective optimization problem into a composite objective function by assigning different weights to each objective. It is important to note that time cost and distance are valued on different scales, so they cannot be simply added together. The process of converting the AG-MAFT problem into a single-objective optimization problem through the linear weighting method involves two steps: first, scaling the original objective functions, and second, determining the weights. The scaling model we use is shown in Eq. \ref{equ-13}:
\begin{equation}
    f'(x) = \frac{f(x) - f^{min}}{f^{max} - f^{min}}
    \label{equ-13}
\end{equation}
In this model, $f(x)$ represents the value of the objective function, while $f^{max}$ and $f^{min}$ are respectively represent the maximum and minimum values of the objective function.

Since the objectives of move time cost and travel distance belong to different dimensions, we use two weights, $k_t$ and $k_d$, representing the weights for move time cost and travel distance, respectively. By applying weighted summation to the normalized objective functions, we formulate a single-objective optimization problem. Consequently, the objective function for AG-MAFT is transformed into Eq. \ref{equ-14}:

\begin{equation}
\begin{split}
\text{minimize} \quad & k_t \times \frac{\sum_i T_i \cdot \sum_j x_{ij} - T_{min}}{T_{max} - T_{min}} \\
& + k_d \times \frac{\sum_i \sum_j D_{ij} \cdot x_{ij} - D_{min}}{D_{max} - D_{min}}   
\end{split}
\label{equ-14}
\end{equation}

Clearly, the AG-MAFT problem is an Integer Linear Programming (ILP) problem, and we use the branch-and-bound method for ILP to obtain the optimal solution for the weights. Based on the above analysis, we also apply the K-means clustering algorithm for preliminary task classification.

\begin{algorithm}[!h]
    \caption{W-ILP}
    \renewcommand{\algorithmicrequire}{\textbf{Input:}}
    \renewcommand{\algorithmicensure}{\textbf{Output:}}
    \begin{algorithmic}[1]
    \REQUIRE The agent set \( A(M) \) and the task set \( T(N) \).  
    \ENSURE Agent-task allocation and minimum total distance and time cost.    
    \STATE \textbf{Identify Task-Intensive Regions:} Define task execution regions and assign existing agents with capabilities and communication abilities.
    \STATE \textbf{Agent-Task Matching:} Minimize the weighted sum of time and distance for assignments:
    \[
    \text{minimize} \quad \sum_{i,j} \left( k_t \cdot \text{time}_{ij} + k_d \cdot \text{distance}_{ij} \right) x_{ij}
    \]
    subject to:
    \[
    \sum_{j} x_{ij} = p_i, \quad \sum_{i} x_{ij} \leq 1, \quad x_{ij} \in \{0, 1\}
    \]
    After assignment, UAVs that receive tasks can move toward the task locations.
    \STATE \textbf{Establish UGV-UAV Sets:} Group UAVs around UGVs using K-means or nearest-distance methods. Assign each UAV to a UGV for recharging.
    \STATE \textbf{Return UAVs for Charging:} After tasks are completed, UAVs return to UGVs for recharging. 
    \end{algorithmic}
\end{algorithm}
The W-ILP algorithm is used to solve the AG-MAFT problem and consists of two parts: the first part calculates the maximum and minimum values of time and distance, and applies the linear weighting method to generalize the new single-objective optimization problem; the second part uses the branch-and-bound method to solve this problem. The W-ILP algorithm greedily selects agents with the shortest travel time, meaning agents with faster movement speeds are chosen to perform tasks to minimize move time, resulting in a potential increase in total distance traveled. Additionally, when considering total travel distance, each task should be performed by the closest agent to minimize overall distance, disregarding time cost.

Through these steps, the AG-MAFT problem is successfully transformed into a single-objective optimization problem, and further into an integer linear programming problem, where the optimal solution is obtained via the branch-and-bound method. W-ILP first generates a rooted tree based on the relaxed solution of the ILP problem, determining upper and lower bounds for each current solution. Subsequently, the original optimal solution is continuously adjusted until the integer constraints for the variables are satisfied, involving three steps: branching, bounding, and pruning. Finally, W-ILP outputs the number of agents per task in each area.

To improve task allocation efficiency, this paper also applies the K-means clustering algorithm for preliminary task classification. The K-means clustering algorithm groups tasks into clusters based on the location information of UAVs and UGVs, so that tasks within each cluster are geographically close, thereby reducing the travel distance for agents.

\subsubsection{UAV Charging Path Optimization}
In the current fields of urban management and environmental monitoring, as the demand for refined and real-time data keeps growing, UAVs have emerged as powerful tools for long-term tasks like urban security patrols and air quality monitoring, thanks to their flexible mobility and high-altitude perspective. However, the limited battery capacity of UAVs struggles to meet the demands of uninterrupted long-term operations. In the AG-MAFT scenarios, where such long-term tasks are frequently carried out, the charging requirements of UAVs are especially critical. Once the UAV battery runs out during a mission, it not only interrupts the monitoring data flow, undermining the real-time control of the city, but also may trigger safety hazards due to unexpected landings. Consequently, it is urgent to delve into the UAV charging issue and devise an efficient and suitable charging mechanism to ensure rapid power restoration for UAVs during task execution, thus guaranteeing their continuous operation and the accuracy of real-time monitoring. 

In this situation, using UGVs to provide charging support for multiple UAVs is particularly suitable. Therefore, we designed a Predictive Charging Trajectory Planning (PCTP) algorithm, where each UGV is responsible for charging multiple UAVs \cite{wang2022task}, to improve resource utilization and efficiency. This centralized charging management strategy not only reduces the charging time for each UAV but also ensures they can quickly return to their tasks. More importantly, UAVs with low battery levels can autonomously determine the timing and path for charging, a self-decision capability that reduces interruptions in task execution and improves overall operational efficiency.

The PCTP algorithm calculates the energy consumption required for a UAV to move to its task location based on the UAV’s current position and task location. This energy consumption is proportional to the travel distance, flight time, and task execution time. Since each UGV in a given area is responsible for charging multiple UAVs, UAVs must autonomously move to the UGV’s location for charging. When a UAV's battery reaches a critical level during task execution, it communicates with the UGV to calculate their distance and reserves enough energy to fly back to the UGV for charging.

\begin{figure}[htbp]
    \centering
    \includegraphics[width=0.75\linewidth]{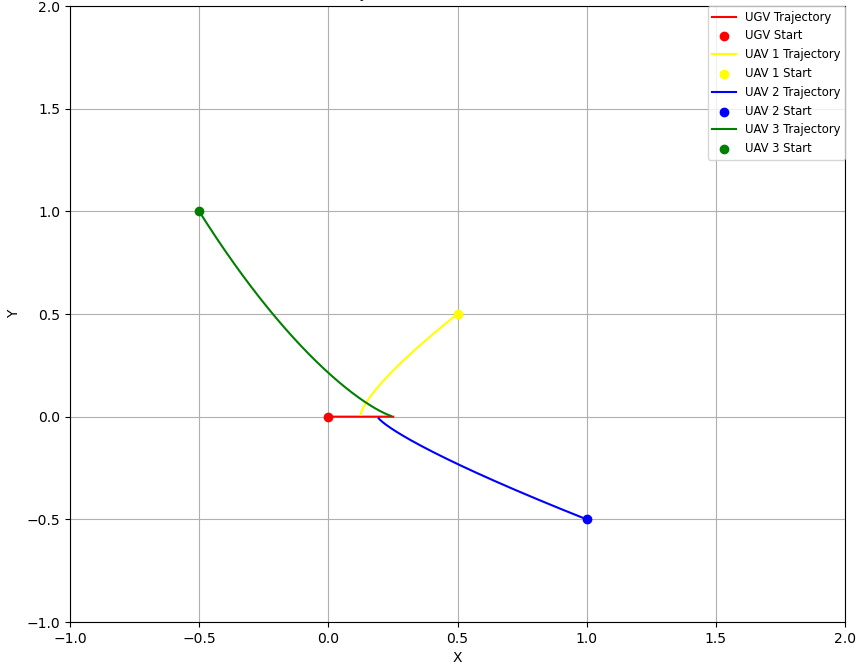}
    \vspace{-0.5em}
    \caption{Charging trajectory of UAVs flying towards UGV.}
    \label{fig3}
\end{figure}

During the period when the UAV is approaching the UGV for charging, the UGV calculates and adjusts its direction and speed based on the movement speed and trajectory of the UAV it is responsible for. Specifically, the direction of movement of the UGV is determined by the vector sum of coordinates between its current position and the UAVs flying towards it for charging. The movement speed in each direction is determined by the flight speed of the UAV, the remaining battery, and the initial distance. The velocity components of the UGV in each direction are inversely proportional to the flight speed $v_{UAV_i}$ of the UAV and the remaining battery $E_{UAV_i}$, and directly proportional to the initial distance $d_{UAV_i}$. The final movement speed of the UGV is the sum of the velocity vectors in these directions. The UGV moves in the calculated direction and at the designated speed, while the UAV approaches the UGV until the distance between them is less than the preset charging distance, at which point charging begins. The trajectory of the UAV charging the UGV is shown in Figure \ref{fig3}.

The initial position of the $UGV$ is $(X, Y)$, and the initial position of each $UAV_i$ is $(x_i, y_i)$. The movement direction $\mathbf{D}_{UGV}$ of the $UGV$ is determined by the sum of the position vectors between it and each $UAV_i$:

\begin{equation}
    \mathbf{D}_{UGV} = \sum_{i} \left( (x_i - X), (y_i - Y) \right)
\end{equation}

Then, $\mathbf{D}_{UGV}$ is normalized to obtain the unit direction vector $\hat{\mathbf{D}}_{UGV}$:

\begin{equation}
\begin{split}
        \hat{\mathbf{D}}_{UGV} &= \frac{\mathbf{D}_{UGV}}{\|\mathbf{D}_{UGV}\|} \\
        &= \frac{\sum_{i} \left( (x_i - X), (y_i - Y) \right)}{\sqrt{\left( \sum_{i} (x_i - X) \right)^2 + \left( \sum_{i} (y_i - Y) \right)^2 }}
\end{split}
\end{equation}

UGV Speed Calculation: Let each $UAV_i$ have a flight speed of $v_{UAV_i}$, remaining battery level $E_{UAV_i}$, and initial distance to the $UGV$ denoted by $d_{UAV_i}$, which can be expressed as:
\begin{equation}
   d_{UAV_i} = \sqrt{(x_i - X)^2 + (y_i - Y)^2} 
\end{equation}

The speed component of the $UGV$ in each direction, $v_{UGV_i}$, is inversely proportional to $v_{UAV_i}$ and $E_{UAV_i}$, and directly proportional to $d_{UAV_i}$, which can be expressed as:
\begin{equation}
\begin{split}
        v_{UGV_i} &= k \cdot \frac{d_{UAV_i}}{v_{UAV_i} \cdot E_{UAV_i}} \\
        &= k \cdot \frac{\sqrt{(x_i - X)^2 + (y_i - Y)^2}}{v_{UAV_i} \cdot E_{UAV_i}}
\end{split}
\end{equation}

Here, $k$ is a proportional constant.

Finally, the $UGV$’s final velocity vector $\mathbf{v}_{UGV}$ is:
\begin{equation}
    \mathbf{v}_{UGV} = v_{UGV} \cdot \hat{\mathbf{D}}_{UGV} = \left( \sum{i} v_{UGV_i} \right) \cdot \hat{\mathbf{D}}_{UGV}
\end{equation}

The preset charging distance $D$ is:
\begin{equation}
    \sqrt{(x_i - X)^2 + (y_i - Y)^2 }\ \leq \ D,
\end{equation}

In comparison, in the AG-FAMT scenario, tasks are generally urgent, so the charging demand for UAVs is relatively lower for time-sensitive tasks. Using the MT-MCMF model for task allocation, each agent receives a task set with a number of tasks up to its maximum executable task limit $q_i$, and the travel distance required to complete the task set is within its maximum allowable distance $S$. In this context, agents can effectively manage power consumption while performing urgent tasks through optimized path planning and task allocation, reducing the need for frequent charging.

In summary, by designing a Predictive Charging Trajectory Planning algorithm for UAVs flying to UGVs, we can ensure the continuous and efficient operation of UAVs in AG-MAFT scenarios, enhancing their reliability and efficiency in long-term monitoring tasks. This optimization strategy not only improves task execution stability but also provides valuable insights for future UAV applications.

\section{EVALUATION}
\subsection{Dataset and Experiment Setups}
\subsubsection{Dataset}

In the simulation scenario, we consider a signal cell consisting of multiple UAVs, multiple UGVs and a base station \cite{zhou2019computation}. Given the common problem of data scarcity in ${MCS}$ platforms and the lack of standard baseline methods and dedicated datasets for the research of task allocation methods, we carry out relevant work relying on the D4D dataset \cite{blondel2012data}. In order to train and accurately evaluate the model, we sample from the D4D dataset of real trajectories to initialize the spatiotemporal distribution of the set of sensing tasks. Since the node distribution in this dataset is identified by latitude and longitude, during this process, we convert the latitude and longitude information into two-dimensional coordinates to simulate the task coordinates. Then, we design experiments based on the geographical locations of the cell towers and mobile phone users. At the same time, we set the location-related task information according to the positions of the cell towers, and set the location of the agent as the position of the cell tower where the mobile user makes a call.  

\subsubsection{Experiment Setups}
For AG-FAMT, the maximum number of agents required to complete a task is \( p \). Due to the limited resources of agents and the upper limit of their executable tasks, in order to execute as many tasks as possible and avoid excessive repetition of the same task, perception data redundancy may occur. Therefore, considering that the value of $p$ has no impact on the experimental results, we will set it to 6 in the following experiment. In addition, considering the completion status of tasks, the number of tasks that each agent can perform at once is randomly set between 2 and 7. To assess the completion time of an agent with known movement distances, we set the average moving speed of a UGV at 5 meters per minute and the average moving speed of a UAV at 20 meters per minute. In addition, to evaluate whether our proposed method performs well under different task space distributions, we used three types of task distributions.

- Compact Distribution: Tasks are relatively close together and are mainly concentrated within the region where agents are located.

- Scattered Distribution: Tasks are scattered across a broader target area beyond the region where agents are located.

- Hybrid Distribution: Tasks are randomly distributed across the region where agents are located.

\begin{figure}[htbp]
    \centering
    \includegraphics[width=1\linewidth]{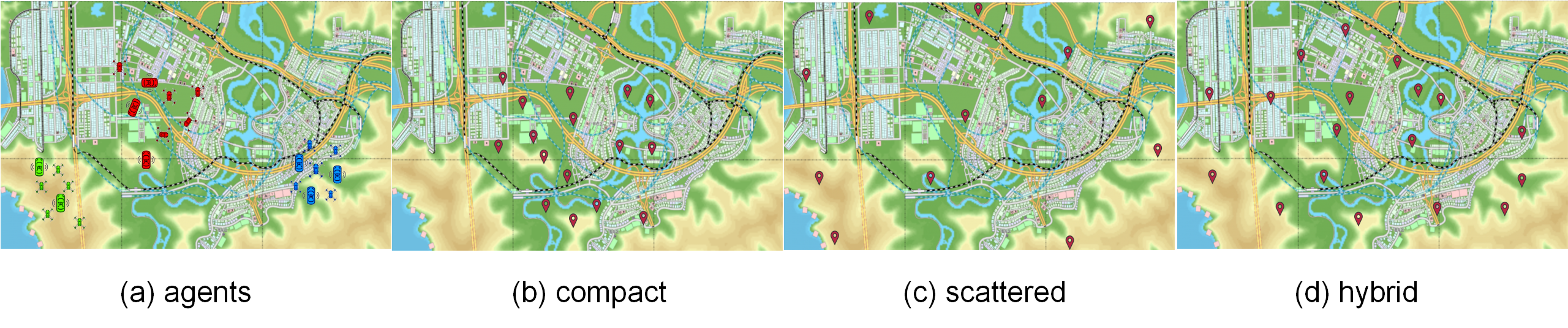}
    \vspace{-0.5em}
    \caption{An illustration of the three types of task distribution.}
    \label{fig4}
\end{figure}

An illustration of the three types of distributions used in our experiments is given in Figure \ref{fig4}.

\begin{figure}[htbp]
    \centering
    \includegraphics[width=1\linewidth]{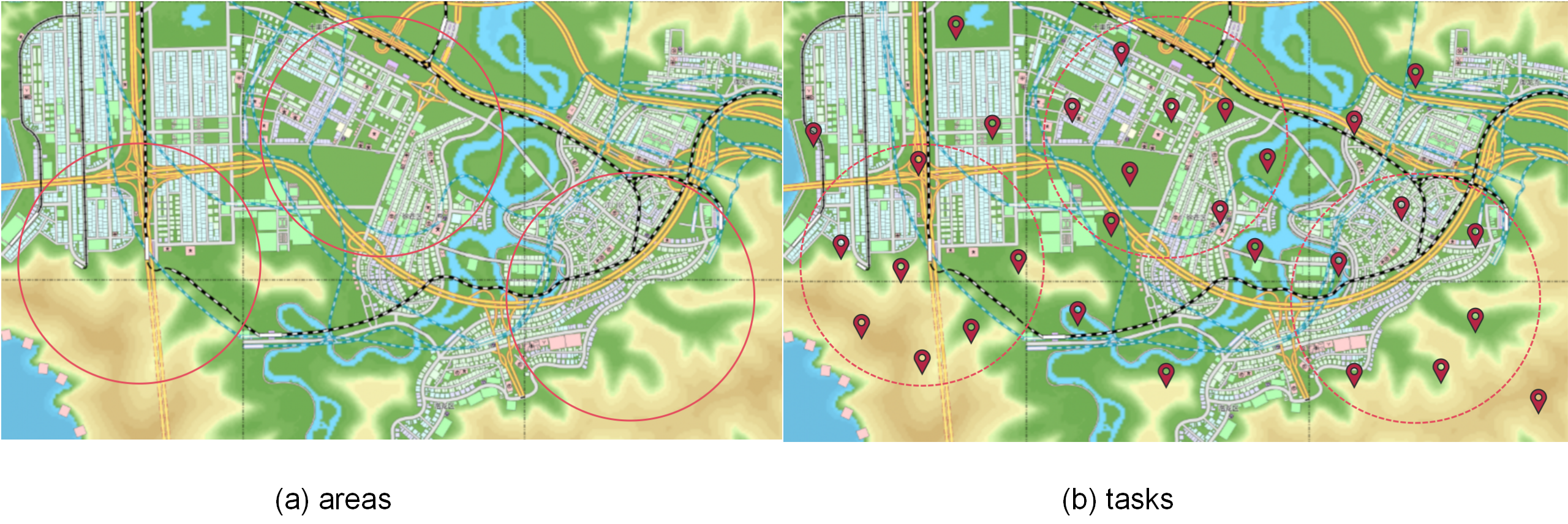}
    \vspace{-0.5em}
    \caption{Distributions of areas and tasks.}
    \label{fig5}
\end{figure}

For AG-MAFT, to protect privacy, each agent’s precise location is unknown before task execution and is instead represented by the area’s location.  In our experiment, agents across the entire area are divided into 3 to 5 sections, evenly distributed throughout the city. The number of agents in each area is randomly generated between 5 and 20.  Additionally, for the limited-task AG-MAFT scenario, we assume there are 20 randomly distributed tasks throughout the city, each requiring 5 agents to ensure sensing quality.  This setup is illustrated in Figure \ref{fig5}.

\subsection{Baseline Methods and Evaluation Metrics}

In our evaluation, we selected commonly used baseline methods from previous experiments under similar scenarios. We use MT-GrdPT \cite{liu2016taskme} as a baseline method for comparison with AG-FAMT. Additionally, we introduce a baseline method named W-Grd \cite{liu2016taskme} for AG-MAFT and a static UGV charging algorithm \cite{7822448},\cite{10.1115/IMECE2012-88246} as a baseline for comparison with PCTP.

\begin{itemize}
    \item MT-GrdPT: The goal of the MT-GrdPT algorithm is to minimize travel distance while maximizing the number of completed tasks. It uses a greedy approach where each agent selects tasks based on the shortest distance to a task until the agent completes $q$ tasks. It is important to note that each agent must move from its current position to the assigned task locations. 
    \item W-Grd: The W-Grd algorithm is a greedy algorithm based on the linear weighting method, converting a multi-objective problem into a single-objective problem. Specifically, this method applies the linear weighting approach to transform multi-objective optimization (e.g., minimizing total travel time and total travel distance) into a single-objective optimization. It assigns weights to each objective function and forms a composite objective function through weighted summation.
    \item Static-UGV charging: In this model, the UGV remains stationary, and each UGV gradually approaches the UGV at a constant speed until the distance between them reaches a preset charging distance, simulating the charging process upon arrival at the UGV.
\end{itemize}

\subsection{Evaluation Metrics}
For AG-FAMT, the total number of tasks completed by the selected agents and the total travel distance are the primary comparison metrics. Additionally, for some urgent tasks in the AG-FAMT problem, the average completion time for $q$ tasks should be as short as possible. Lastly, the algorithm's runtime is also an important factor in solving the problem.

For AG-MAFT, the total travel distance covered by agents during task completion is a key comparison metric, while the cost of time taken to reach the task execution point is another important consideration. Additionally, in the AG-MAFT scenario, the total travel distance of agents during the UAV charging process is also a primary comparison metric.

\subsection{Evaluation of AG-FAMT}
The optimization objective of AG-FAMT is to minimize the total travel distance while maximizing the total number of completed tasks. It is important to note that, since $M$ agents are required to perform $q$ tasks each, the total number of tasks completed by the agents is $M \times q$. We compare the performance (in terms of total travel distance) of the two algorithms under different total task quantities, different numbers of agents, different task requirements per agent $q$, and different types of task distributions.

\subsubsection{Different Numbers of Tasks}

\begin{figure}[htbp]
    \centering
    \includegraphics[width=0.75\linewidth]{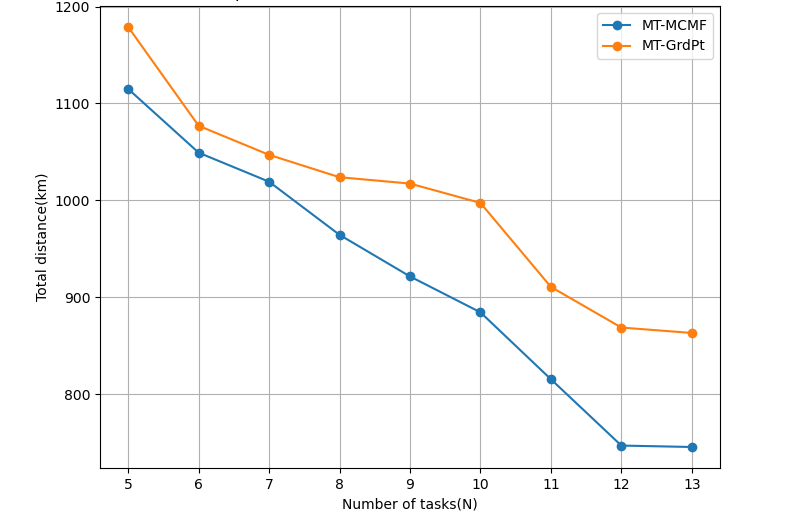}
    \vspace{-0.5em}
    \caption{Performance comparison under different number of tasks.}
    \label{fig6}
\end{figure}

As shown in Figure \ref{fig6}, we present the performance comparison of total travel distance for the two methods under different total task quantities. Additionally, we assume that there are 4 agents on the MCS platform performing sensing tasks, with each agent required to complete 3 tasks. As shown in Figure \ref{fig6}, when the four agents execute tasks, the total travel distance remains constant or decreases monotonically as the total number of tasks increases. This is because additional tasks may provide better task options for the agents. Clearly, MT-MCMF outperforms MT-GrdPT across all task quantities.

\subsubsection{Different Numbers of Agents}

\begin{figure}[htbp]
    \centering
    \includegraphics[width=1\linewidth]{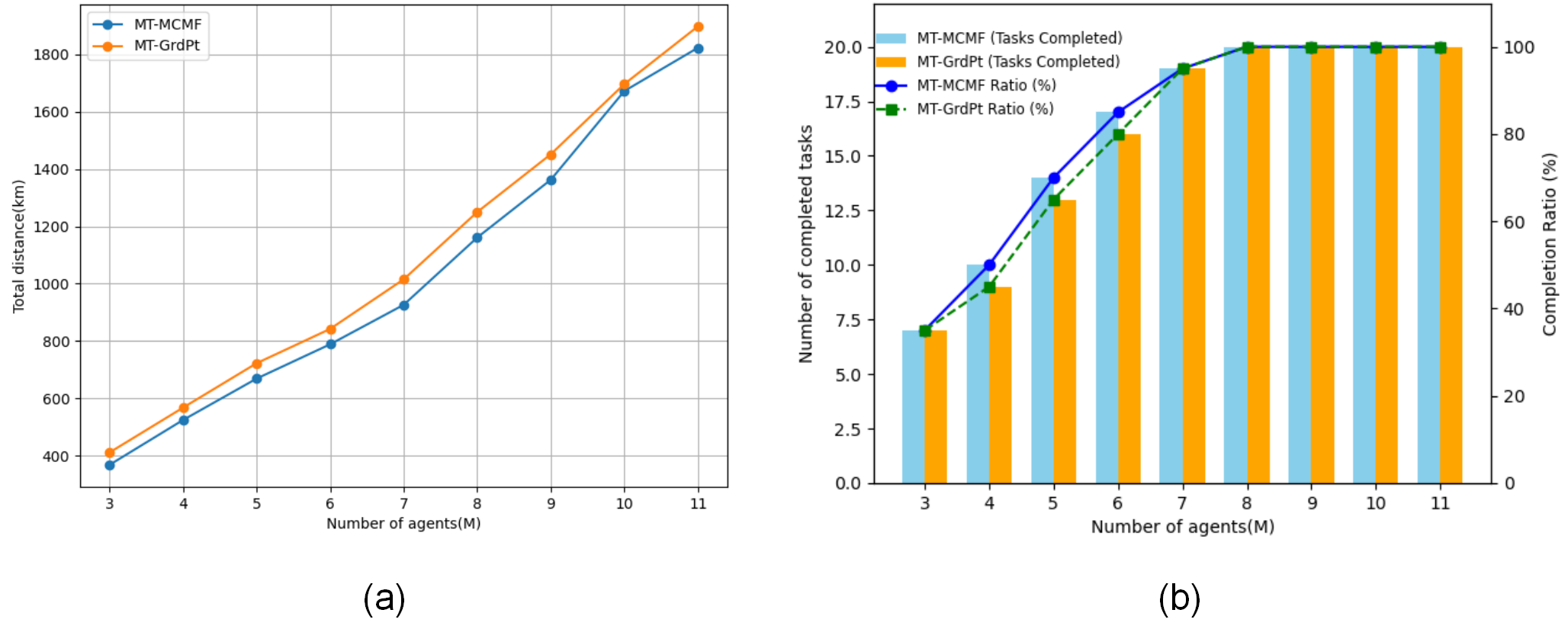}
    \vspace{-0.5em}
    \caption{Performance comparison under different number of Agents.}
    \label{fig7}
\end{figure}

As illustrated in Figure \ref{fig7}, we compare the total travel distance of two methods under varying numbers of agents for the same number of tasks. Note that the number of agents $M$ does not exceed the number of tasks $N$. There are 20 sensing tasks on the MCS platform, with the number of agents ranging from 3 to 11, and each agent is required to complete 3 tasks. Figure \ref{fig7}(a) shows that as the number of agents increases, the total number of completed tasks $M \times q$ also increases, leading to an increase in total travel distance. And Figure \ref{fig7}(b) illustrates that when each agent needs to complete 3 tasks, if the number of selected agents exceeds one-third of the task count, agents can approximately complete all tasks. Clearly, MT-MCMF outperforms MT-GrdPT in terms of travel distance across different numbers of agents.

\subsubsection{Different Values of q}

\begin{figure}[htbp]
    \centering
    \includegraphics[width=1\linewidth]{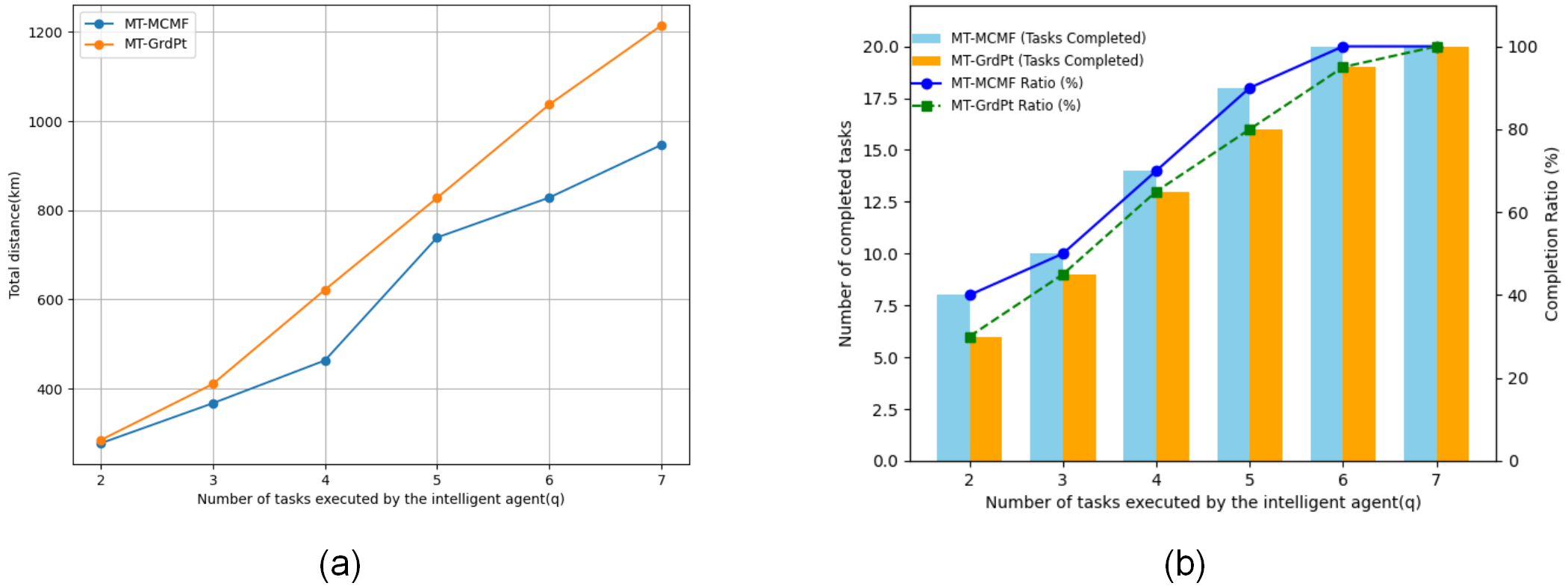}
    \vspace{-0.5em}
    \caption{Performance comparison under different number of q.}
    \label{fig8}
\end{figure}

We assume there are 4 agents available to perform tasks, with each agent required to complete $q$ tasks. We compare the task allocation performance of the two algorithms with different number of tasks for each agent. In Figure \ref{fig8}(a) , as the number of completed tasks $M \times q$ continuously increases, the total number of possible tasks to be executed, $N \times q$, also increases, resulting in a corresponding increase in total travel distance. It is evident that the MT-GrdPT algorithm performs poorly in terms of travel distance, and this discrepancy becomes more pronounced as $q$ increases. MT-GrdPT is a greedy algorithm that tends to reach a local optimum in a relatively short time, leading to suboptimal results as the number of iterations increases. In addition, Figure \ref{fig8}(b) illustrated that as the total number of tasks remains constant, the number of completed tasks is proportional to $q$. When the number of completed tasks $M \times q$ approaches the total number of tasks N, agents are able to complete nearly all tasks.

\subsubsection{Different Types of Task Distributions}

\begin{figure}[htbp]
    \centering
    \includegraphics[width=0.75\linewidth]{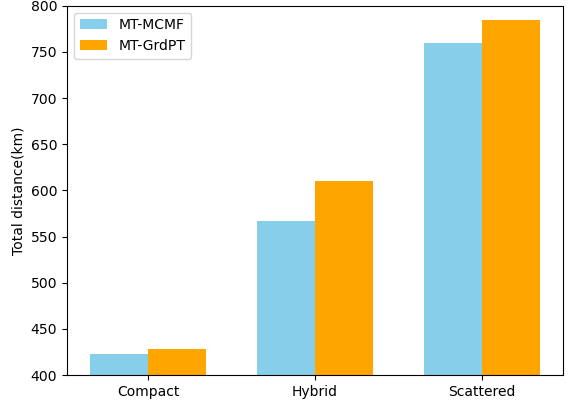}
    \vspace{-0.5em}
    \caption{Different types of Task Distributions.}
    \label{fig9}
\end{figure}

We investigate which type of task distribution is more advantageous for selecting agents to complete tasks: compact, mixed, or dispersed distribution. As shown in Figure \ref{fig9}, we can observe that the dispersed distribution results in the longest total travel distance for agents to complete tasks, while the compact distribution has the shortest distance. This is because tasks in a dispersed distribution are farther apart compared to the other two distributions, requiring agents to travel longer distances and spend more time completing all tasks. Specifically, MT-MCMF outperforms MT-GrdPT across all task distributions.

\subsection{Evaluation of AG-MAFT}
The optimization objective of AG-MAFT is to minimize the total move time cost and total travel distance while ensuring all tasks are completed. For the bi-objective optimization problem, there are some groups of non-inferior solutions instead of the global optimal solution, known as the Pareto solution. Therefore, we examine the results of the proposed algorithm in terms of both total time cost and total travel distance.

\subsubsection{Different Weights}

\begin{figure}[htbp]
    \centering
    \includegraphics[width=1\linewidth]{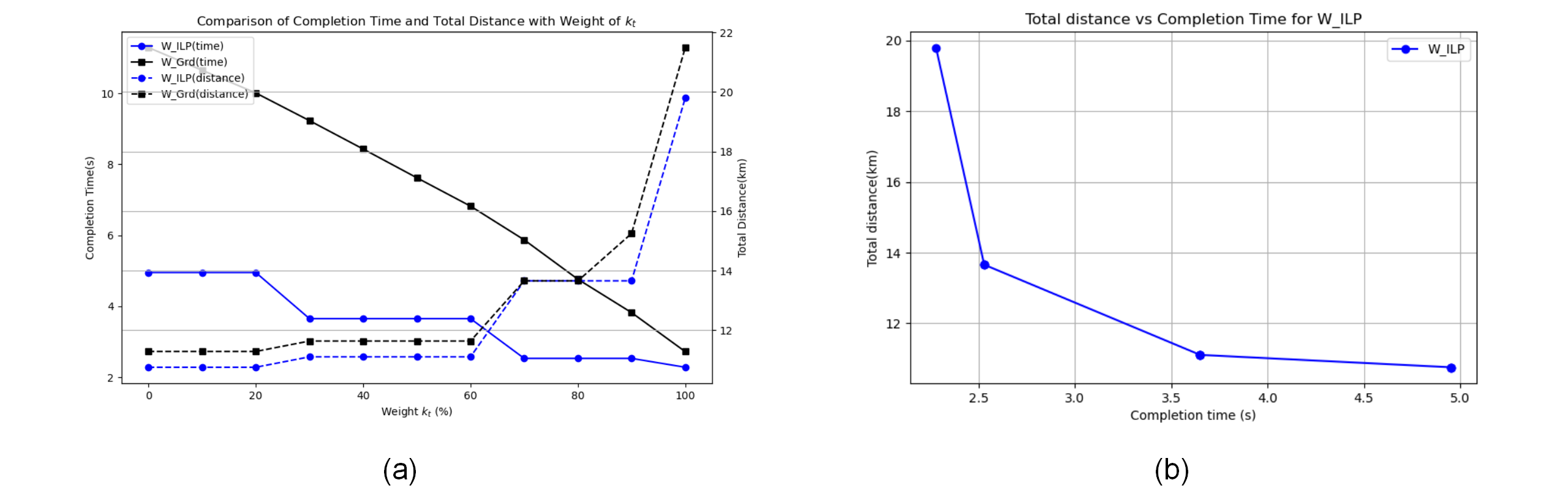}
    \vspace{-0.5em}
    \caption{Different weights of AG-MAFT.}
    \label{fig10}
\end{figure}

Figure \ref{fig10} shows the total travel distance and time of the W-ILP (Integer Linear Programming based on Linear Weighting Method) algorithm and the W-Grd (Greedy Algorithm based on Linear Weighting Method) algorithm under different weights. $K_d$ represents the weight for travel distance, and $K_t$ represents the weight for time, with $K_d$ and $K_t$ clearly summing to 1. As shown in Figure \ref{fig10}(a), with an increase in the weight of $K_t$, the travel time gradually decreases, while an increase in the weight of $K_d$ leads to an increase in travel distance. Consequently, it is preferable to assign farther tasks to faster-moving UAVs, while UGVs execute tasks nearby. Notably, the travel distance results calculated by the W-Grd algorithm are very close to those of the W-ILP algorithm, indicating that the performance of the W-Grd algorithm approaches the optimal solution. Additionally, the W-Grd algorithm also achieves good results in terms of travel time.

According to Figure \ref{fig10}(b), the $Pareto$ solutions of W-ILP can be obtained. It is evident that no single weight value can minimize both travel time and travel distance simultaneously. From a time perspective, the first set of non-dominated solutions is a relatively better choice, while the last set of non-dominated solutions is a relatively better choice for distance. Generally, the third set of non-dominated solutions provides a balanced solution for both time and distance. These results suggest that determining the weights $K_t$ and $K_d$ requires a trade-off between time and travel distance.

The advantage of W-ILP lies in its dual-objective function (i.e., travel distance and travel time), which can be flexibly adjusted according to the needs of task issuers in different scenarios. For example, when task urgency is high or travel distances are long, the weight of travel time $K_t$ should be increased accordingly. However, it is often challenging to set appropriate weight values to ensure solution quality, as even slight differences between the two weight values can significantly impact performance.

\subsubsection{Different Charging Path Selections}

\begin{figure}[htbp]
    \centering
    \includegraphics[width=0.75\linewidth]{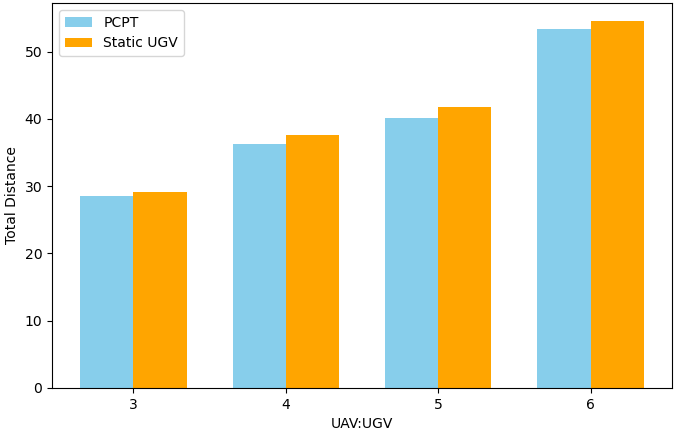}
    \vspace{-0.5em}
    \caption{Different total distance of charging.}
    \label{fig11}
\end{figure}

In the experiment, we divided the setup into four regions and compared charging paths in each region. As illustrated in Figure \ref{fig11}, when a single $UGV$ is responsible for charging different numbers of $UAV$s, the travel distance of the $PCTP$ algorithm is shorter than that of the Static-UGV charging algorithm. This indicates that $PCTP$ can effectively reduce the total travel distance of agents when multiple $UAV$s require charging. This result confirms the optimization effectiveness of $PCTP$ in path planning, demonstrating that it can more efficiently respond to the dynamic demands of $UAV$s through predictive trajectory planning. This leads to effective control of agent travel distance, helping to improve overall system efficiency.

\section{DISCUSSION}

\subsection{Superiority of Air-Ground Collaborative MCS}

This paper proposes a multi-agent collaborative task allocation framework to address the AG-FAMT and AG-MAFT problems. For the AG-FAMT scenario, we introduce a Minimum-Cost Maximum-Flow (MCMF) model to ensure that each homogeneous agent completes an equal number of sensing tasks, thereby maximizing the total number of completed tasks while minimizing movement distances. In the AG-MAFT scenario, we flexibly select suitable agents based on the task demands across different regions and agents’ capabilities, optimizing task allocation outcomes. The framework AGCo-MATA, grounded in quantitative considerations, demonstrates high flexibility across various constraints, offering a novel direction for research in multi-agent collaborative task allocation.

\subsection{Configuration Optimization}
The experimental results demonstrate that the proposed method effectively enhances task completion efficiency in both AG-FAMT and AG-MAFT scenarios. In the AG-FAMT scenario, optimizing the configuration of agent numbers and the number of tasks assigned to each agent maximizes task completion rates while minimizing movement distances. For the AG-MAFT scenario, selecting agents based on time cost and path distance significantly improves task completion rates and system robustness. Additionally, in AG-MAFT, an increase in the weight of \( K_t \) leads to a gradual decrease in travel time, while a higher weight for \( K_d \) increases travel distance. This suggests that assigning farther tasks to faster-moving UAVs and closer tasks to UGVs can further optimize task allocation. These findings indicate that the proposed allocation strategy has broad applicability across diverse application scenarios.
\subsection{Limitations}
Although the research findings provide valuable insights, there are still certain limitations. Firstly, the experiment assumes that the performance of all homogeneous agents is consistent and the difficulty of the task is uniform. In practical applications, these factors often have high heterogeneity and may affect the efficiency of task allocation. Secondly, this study does not consider communication costs or collaboration methods among agents, which may affect overall task completion in real-world applications due to communication delays or coordination mechanisms. Additionally, the task allocation rules in the experiment are relatively idealized and do not fully account for dynamic changes in tasks and environmental complexity, which may affect the applicability of results in dynamic environments.
\subsection{Future Work}
To further enhance task allocation efficiency in the AG-FAMT problem, future research could explore incorporating task prioritization, inter-agent dynamic collaboration mechanisms, and failed task redistribution to refine allocation strategies. In addition, considering heterogeneity in tasks and agents (such as priority of tasks and varying agent capabilities) would help build more precise allocation models. In the AG-MAFT scenario, methods to minimize redundant collaborations among agents are also worth exploring to reduce system resource waste.

\section{CONCLUSION}
In this paper, we proposed two bi-objective optimization problems for agent selection: the AG-FAMT problem, where agent resources are insufficient, and the AG-MAFT problem, where agent resources are abundant. We use the MCMF model to address the AG-FAMT problem and propose the MT-MCMF algorithm to select agents with the maximum number of completed tasks and the minimum total travel distance. Additionally, we employ a dual-objective optimization model to solve the AG-MAFT problem and introduce the W-ILP algorithm to select agents based on travel time and minimum total travel distance. Furthermore, we propose the PCTP algorithm for UAVs charging path planning to optimize the travel distance for agent charging.

\bibliographystyle{IEEEtran}
\bibliography{reff.bib}


 




\vfill

\end{document}